\documentclass[conference]{IEEEtran}
\pdfoutput=1
\IEEEoverridecommandlockouts %To add sponsors and index terms

\usepackage{cite}
\usepackage{acronym}
\usepackage[utf8]{inputenc}
\usepackage[T1]{fontenc}
\usepackage{mathdots}

\renewcommand{\vec}[1]{\ensuremath{\boldsymbol{#1}}} % bold vectors

\ifCLASSINFOpdf
   \usepackage[pdftex]{graphicx}
   \DeclareGraphicsExtensions{.pdf,.jpeg,.png}
\else
\fi
\usepackage[cmex10]{amsmath}
\usepackage{multirow}
\usepackage{booktabs,colortbl}

\begin{document}
\title{Battery Degradation Maps for Power System Optimization and as a Benchmark Reference}

\author{%

\IEEEauthorblockN{Philipp Fortenbacher \\ Göran Andersson }
\IEEEauthorblockA{Power Systems Laboratory\\
ETH Zurich\\
Zurich, Switzerland\\
\{fortenbacher,andersson\}@eeh.ee.ethz.ch}


}

\acrodef{LV}[LV]{Low Voltage}
\acrodef{AC-OPF}[AC-OPF]{AC Optimal Power Flow}
\acrodef{OPF}[OPF]{Optimal Power Flow}
\acrodef{FBS-OPF}[FBS-OPF]{Forward Backward Sweep Optimal Power Flow}
\acrodef{FBS}[FBS]{Forward Backward Sweep}
\acrodef{IP}[IP]{Interior Point}
\acrodef{LP}[LP]{Linear Programming}
\acrodef{SOCP}[SOCP]{Second Order Cone Programming}
\acrodef{SDP}[SDP]{Semi Definite Programming}
\acrodef{SoC}[SoC]{State of Charge}
\acrodef{SoE}[SoE]{State of Energy}
\acrodef{DoD}[DoD]{Depth of Discharge}
\acrodef{PWA}[PWA]{piecewise affine}
\acrodef{ID}[ID]{Identification}
\acrodef{SEI}[SEI]{Surface Electrolyte Interface}
\acrodef{LS}[LS]{Least Squares}

\maketitle

\begin{abstract}
This paper presents a novel method to describe battery degradation. We use the concept of degradation maps to model the incremental charge capacity loss as a function of discrete battery control actions and state of charge. The maps can be scaled to represent any battery system in size and power. Their convex piece-wise affine representations allow for tractable optimal control formulations and can be used in power system simulations to incorporate battery degradation. The map parameters for different battery technologies are published making them an useful basis to benchmark different battery technologies in case studies.     
\end{abstract}


\section{Introduction}
Battery degradation depends on battery operation and can thus be influenced by the operational management. To enhance economic profitability it is crucial to have suitable and functional degradation models, which can be used in battery applications that take battery life time into account. Degradation processes are very hard to model, and even in the electrochemical domain there does not exist a complete understanding of the phenomenon \cite{Ramadass}. 

There has been significant recent research focused on the development of battery life cycle and calendar life models \cite{Millner2010,Wang2014,Xu2014}. These models are semi-empirical that fit experimental data to analytic functions that take as inputs Ah or cycle throughput, \ac{DoD}, temperature, current rates and calculate the remaining charge capacity. However, such representations are not applicable in control applications, since cost functions are needed in such a framework that map an individual control action to an incremental capacity loss. Moreover, these models are also not able to account for arbitrary usage pattern that are present in power system applications.

The main contribution of this paper is the development of degradation maps derived from various sources for Li-ion batteries comprising different cathode materials. By means of degradation maps we can associate an discrete individual control action to an incremental capacity loss. We show that the maps can be scaled to represent any battery system in size and power. The maps can be generically incorporated as convex \ac{PWA} cost functions into an optimal control framework to achieve tractable formulations of the control problem. Also they can be used for carrying out case studies that assess the impact on battery degradation for different battery technologies and different usage pattern. In this paper we publish the map parameters for three battery technologies namely LiCoO2, LiMnNiCo/LiMn2O4, and LiFePO4. It is projected that those battery technologies will have a significant market share in Lithium-based battery applications. In 2025, it is forecasted that these technologies will have a share of 86\% (LiFePO4 (LFP) 26\%, LiCoO2 (LCO) 26\%, LiMn2O4 (LMO) 9\% and LiMnNiCo (NMC) 25\%) \cite{battery_market} of the available Lithium cathode materials.   

\section{Degradation Modeling}
Degradation or capacity fade has various physical phenomena. A comprehensive overview on fading mechanisms is given in \cite{Vetter2005} for Li-ion batteries. According to the authors anode, cathode, and electrolyte aging mechanisms attribute to capacity fade. A more quantitative study on these mechanisms is reported by Arora et al. \cite{Arora1998}, where they identified various processes contributing to degradation. The most relevant are overcharge, and interfacial film formation. All processes can be characterized by chemical side reactions that irreversibly transform cyclable ions into solid components and lead to active material loss and consequently to charge capacity loss. 

In general, chemical side reactions are activated by voltage potentials, temperature, and current rates. Not all side reactions and mechanisms are understood in detail making them a relevant topic for research. Since degradation is very hard to model and can only be formulated properly in the complex electrochemical domain, research also concentrates on finding of macroscopic semi-empirical models. Importantly, only such functional models can be integrated in power system applications that make use of optimization techniques.  

\begin{figure}[!t]
	\centering
	\def\svgwidth{\columnwidth}
	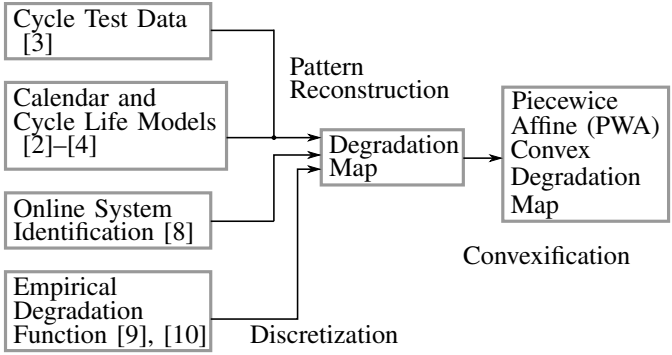
	\caption{Procedure to convert existing degradation models into degradation maps. Those maps associate an incremental capacity loss to an individual control action. For convex optimization the degradation maps have to be convexified or approximated to convex functions.}
	\label{fig:degOverview}
\end{figure}

As shown in Fig.~\ref{fig:degOverview}, a procedure has been developed to convert available models into degradation maps. For their determination three cases are distinguished:
\begin{LaTeXdescription}
	\item [System Identification] An online \ac{ID} method is presented in \cite{fortenbacherPSCC14} that maps the lost charge capacity associated with each discrete control action to the \ac{SoC} and current, allowing to create a battery degradation map. 
	\item [Pattern Reconstruction] Using the aforementioned approach, a battery usage pattern can be reconstructed from either cycle test data or life cycle and calendar life models. 
	\item [Disrectization] Empirical analytic degradation functions as reported in \cite{Forman2012,Li2011} can also be represented in the proposed framework by discretization.
\end{LaTeXdescription}  

Following sections detail the procedure in all steps.  

\subsection{Degradation Maps}
\label{sec:degradmaps}
The concept of degradation maps is not novel. Moura et al. \cite{Moura} used such concept to approximate the film growth rate as a function of the \ac{SoC} and battery current. Since they developed the map to describe only the \ac{SEI} layer decomposition, this work adopts this idea to generalize this approach also describing hidden and unknown chemical side reactions as a black box model. A stationary degradation process is assumed. Side reactions are activated by 1) potential differences between the interfaces of electrolyte and electrodes, 2) temperature $\theta$, and  3) the applied battery current $I_\mathrm{bat}$ \cite{Ramadass}. The reaction rate $I_\mathrm{s}$, defined as the number of ions converted into solid material per time, is related to the side current 
\begin{equation}
I_\mathrm{s} = -\dot{C}_{\mathrm{Q}} = \dot{h}(V_{\mathrm{oc}},I_\mathrm{bat},\theta) \quad, \label{eq:degradation}
\end{equation}
\noindent where $V_{\mathrm{oc}}$ is the open-circuit voltage and is a proxy for the \ac{SoC}. ${C}_{\mathrm{Q}}$ denotes the battery charge capacity. Since this generalized process takes place internally, this current cannot be measured directly. Also, measurement of the internal resistance is not sufficient for determining the complete degradation process because it is affected by only the anode layer decomposition. The aim of the two following sections is to identify the unknown process \eqref{eq:degradation} by using a system identification method, taking calendar and cycle life models from existing literature or cycle test data into account.

\subsection{System Identification}
\label{sec:SIdegrad}
\begin{figure}[!t]
	\centering
	\def\svgwidth{\columnwidth}
	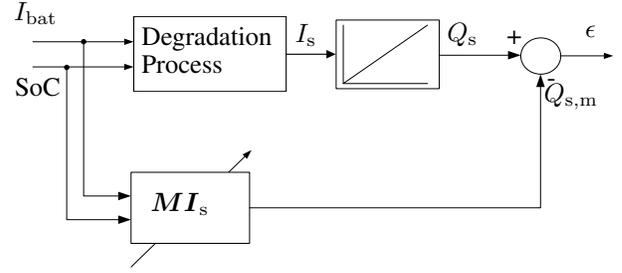
	\caption{Block diagram of the online degradation system \ac{ID} process to obtain degradation maps.}
	\label{fig:degID}
\end{figure}

First, we shortly review the system \ac{ID} method from \cite{fortenbacherPSCC14} to show the extensions for the pattern reconstruction case. The lost charge $Q_\mathrm{s}$ can only be measured over long observation periods \cite{Plett20112319}. Hence, it cannot be directly associated with individual control actions. However, if there are many charge capacity measurements, it is possible to estimate the charge lost from discrete types of control actions and project the lost charge with $\vec{M}_\mathrm{p}$ to the involved usage pattern in a system \ac{ID} process as depicted in Fig.~\ref{fig:degID}.

A control action is defined by the applied battery current and the battery's SoC, and so these two values are discretized into $n_\mathrm{bd}$ SoC `bands' and $m_\mathrm{bd}$ current `intervals'. It is assumed that the side current is a function of these values (note that isothermal battery operation is assumed such that the direct effect of temperature is neglected). For $n_\mathrm{m}$ capacity loss measurements ($Q_{\mathrm{s},1},...,Q_{\mathrm{s},n_\mathrm{m}}$) and arbitrary stimuli patterns, the unknown discretized side current values can be arranged into a system of linear equations 
\begin{equation}
\resizebox{\hsize}{!}{$\underbrace{\left[\begin{array}[h]{c}
	\sum\limits_{j=1}^{m_{\mathrm{bd}}}\sum\limits_{l=1}^{n_{\mathrm{bd}}}p_{1,(n_\mathrm{bd}(j-1)+l)}\, T_{\text{b},j} \, {I}_{\text{s}}\left(\frac{2l-1}{2n_\mathrm{bd}},\bar{I}_{\mathrm{bat},j} \right) \\
	\vdots 	\\
	\sum\limits_{j=1}^{m_{\mathrm{bd}}}\sum\limits_{l=1}^{n_{\mathrm{bd}}}p_{n_\mathrm{m},(n(j-1)+l)} \, T_{\text{b},j} \, {I}_{\text{s}} \left( \frac{2l-1}{2n_\mathrm{bd}},\bar{I}_{\mathrm{bat},j} \right)
	\end{array}\right]}_{\vec{M}_\mathrm{p}{\vec{I}}_{\mathrm{s}}}  = \underbrace{\left[\begin{array}[h]{c}
	Q_{\mathrm{s},1} \\
	\vdots\\
	Q_{\mathrm{s},n_\mathrm{m}} 
	\end{array}\right]}_{\vec{Q}_{\mathrm{s}}} , $}
\end{equation}
where $p$ counts the control actions associated with each \ac{SoC} band and current interval. The variable $T_{\text{b},j}$ is the time for traversing the corresponding \ac{SoC} band. To avoid an underdetermined system, it follows that $n_\mathrm{m} \geq n_\mathrm{bd}m_\mathrm{bd}$. The unknown vector $\vec{I}_{\mathrm{s}}$ can be estimated with \ac{LS} 
\begin{equation}
\begin{split}
{\vec{I}}_{\mathrm{s}} =& \arg\underset{{\vec{I}}_{\mathrm{s}}}{\min} \|\vec{M}{\vec{I}}_{\mathrm{s}} - \vec{Q}_{\mathrm{s}}  \|_2^2 \quad ,  \\ & \mathrm{s.t.} \quad {\vec{I}}_{\mathrm{s}}  > 0  \quad .
\end{split} 
\label{eq:Is}
\end{equation}

Since the resulted degradation map is nonconvex the map has to be further processed to obtain a convex representation that can be used in a convex optimization framework. Note also that $\vec{M}_\mathrm{p}$ has to consist of linearly independent rows, which means full rank, to find an unique solution of $\vec{I}_\mathrm{s}$ in \eqref{eq:Is}. 

\subsection{Pattern Reconstruction for Determining Degradation Maps}
\label{sec:empiricalCycleTest}
The system \ac{ID} approach from previous Section~\ref{sec:SIdegrad} can be used to determine degradation maps for (i) life cycle and calendar life models and (ii) cycle test data. References \cite{Millner2010,Wang2014,Xu2014} represent (i) where the total capacity loss for a battery cell is defined as the sum of life cycle and calendar life wear and is calculated from full cycle achievement, temperature, and current data. Case (ii) could be cycling test data provided by battery manufactures that correlates the full cycle achievement $n_{\mathrm{cyc}}$ with different \acp{DoD}. For case (ii), the involved current pattern can be reconstructed from this information that has caused the capacity loss. For each individual  $\text{DoD}_i$, we need the number of full cycles that are achieved for a resulting capacity loss value ${Q}_{\mathrm{s},i}$. Assuming that the \ac{DoD} swing is centered at an SoC level of 0.5, the following system of linear equations can be derived:     
\begin{equation}
\resizebox{\hsize}{!}{$\underbrace{\left[
	\begin{array}[h]{ccc}
	\vec{0} & T_\mathrm{b} p_1 & \vec{0} \\
	\iddots & \vdots & \ddots \\
	T_\mathrm{b} p_{n_\mathrm{m}} & \hdots & T_\mathrm{b} p_{n_\mathrm{m}}
	\end{array}\right]}_{\vec{M}_\mathrm{p}(T_\mathrm{b})} 
\underbrace{\left[\begin{array}{c} I_\mathrm{s}(\frac{1}{2n_\mathrm{bd}},I_\mathrm{bat}) \\
	\vdots \\
	I_\mathrm{s}(\frac{2k-1}{2n_\mathrm{bd}},I_\mathrm{bat})\\
	\vdots \\
	I_\mathrm{s}(\frac{2n_\mathrm{bd}-1}{2n_\mathrm{bd}},I_\mathrm{bat})
	\end{array} \right]}_{\vec{I}_\mathrm{s}(I_\mathrm{bat})} = \underbrace{\left[\begin{array}[h]{c}
	Q_{\mathrm{s},1} \\
	\vdots\\
	Q_{\mathrm{s},n_\mathrm{m}} 
	\end{array}\right]}_{\vec{Q}_{\mathrm{s}}(I_\mathrm{bat})} ,$}
\end{equation}
\noindent where $n_\mathrm{m} \geq n_\mathrm{bd}$. Note that the matrix $\vec{M}_\mathrm{p}$ is full rank and the pattern number $p_i$ of each SoC band contribution is calculated as follows 
\begin{equation}
\label{eq:p}
p_i = 2\frac{n_{\mathrm{cyc},i}}{\text{DoD}_i} \quad, 
\end{equation}
\noindent and the time to traverse one \ac{SoC} band is
\begin{equation}
\label{eq:Tb}
T_\mathrm{b} =  \frac{C_\mathrm{Q}}{I_{\mathrm{bat}}n_\mathrm{bd}} \quad. 
\end{equation} 
The map can also be generated for different current rates ($I_{\mathrm{bat},1}, \hdots, I_{\mathrm{bat},m_\mathrm{m}}$), where $m_\mathrm{m}$ denotes the number of current rates. By applying
\begin{equation}
\label{eq:pattern}
\resizebox{\hsize}{!}{$\left[\arraycolsep=1pt \begin{array}{ccc}\vec{M}_\mathrm{p}(T_{\mathrm{b},1}) && \vec{0} \\  & \ddots & \\ \vec{0} &  & \vec{M}_\mathrm{p}(T_{\mathrm{b},m_\mathrm{m}}) \end{array}\right] \left[\arraycolsep=1pt\begin{array}{c} \vec{I}_\mathrm{s}(I_{\mathrm{bat},1}) \\ \vdots \\ \vec{I}_\mathrm{s}(I_{\mathrm{bat},m_\mathrm{m}}) \end{array} \right] = \left[\arraycolsep=1pt\begin{array}{c} \vec{Q}_\mathrm{s}(I_{\mathrm{bat},1}) \\ \vdots \\ \vec{Q}_\mathrm{s}(I_{\mathrm{bat},m_\mathrm{m}}) \end{array} \right] $}
\end{equation}
the same degradation map \eqref{eq:Is} can be obtained as stated in Section~\ref{sec:SIdegrad}. The life cycle models for case (i) can be incorporated in the same way. The only difference is that the degradation values need to be replaced by evaluating the corresponding model functions for the given stimuli pattern $\vec{M}_\mathrm{p}$.

\subsection{Empirical Degradation Functions}
\label{sec:empiricalDegFunction}
\begin{figure}[!t]
	\centering
	\includegraphics[width=\columnwidth]{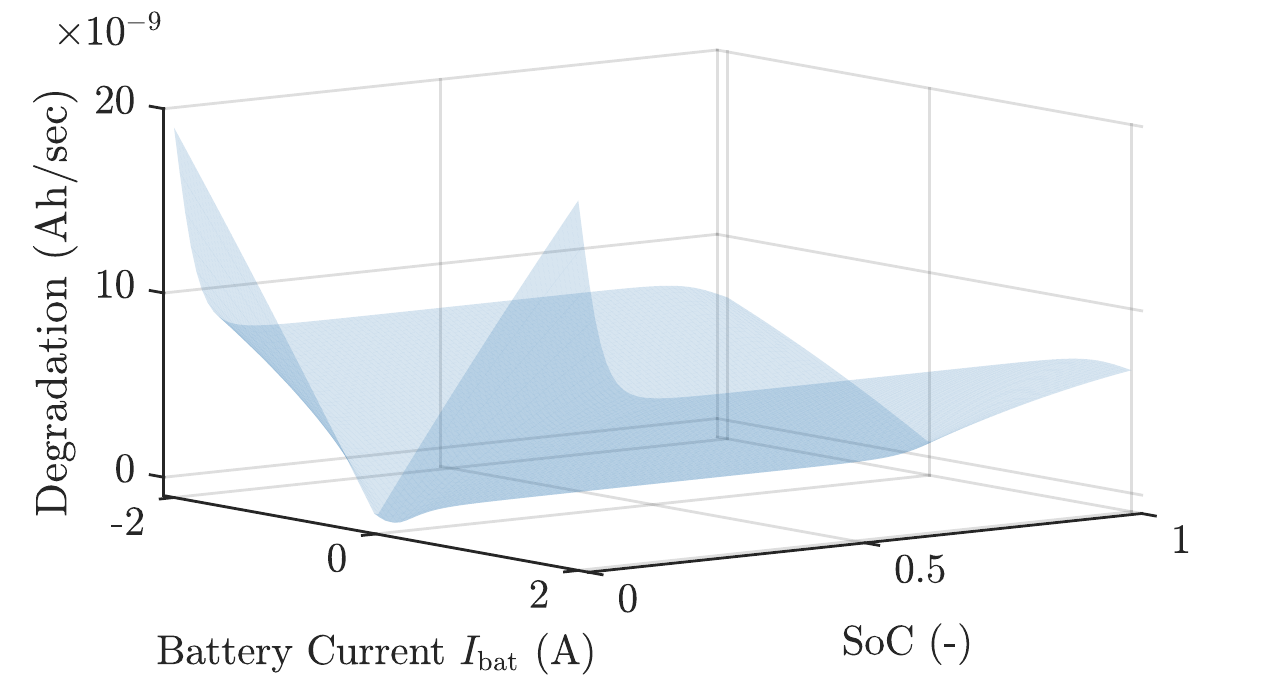}
	\caption{Nonconvex analatytic capacity fade function as a function of the SoC and the battery current from \cite{Forman2012} after SoC substitution.}
	\label{fig:moura_deg}
\end{figure}

The authors in \cite{Li2011,Forman2012} developed empirical degradation functions that already associate degradation with individual control actions. Hence, those approaches represent another form of degradation maps that can also be included in the presented framework. This can be done by discretizing the map to obtain the same representation specified in Sections~\ref{sec:SIdegrad} and \ref{sec:empiricalCycleTest}. As an example, this procedure is presented for the results of \cite{Forman2012} for a LiFePO4 cell. The authors in \cite{Forman2012} derive an analytic expression for capacity loss that maps the control action to capacity loss. In contrast to Sections~\ref{sec:SIdegrad} and \ref{sec:empiricalCycleTest} they use nonlinear regressors for the battery current and the open-circuit potential in the following analytic expression:
\begin{align}
I_\mathrm{s} = \dot{h}(I_\mathrm{bat},V_\mathrm{oc}) = &\beta_1 + \beta_2 |I_\mathrm{bat}| + \beta_3 V_\mathrm{oc} + \beta_4 |I_\mathrm{bat}|^2   \\\nonumber
 & + \beta_5 V_\mathrm{oc}^2 + \beta_6 |I_\mathrm{bat}|V_\mathrm{oc} + \beta_7 V_\mathrm{oc}^3 \quad ,
\end{align}
\noindent where $\beta_1,\ldots,\beta_7$ are fitting parameters. The function $\dot{h}$ describes the capacity loss per time in Ah/sec, is nonconvex and represents the side current. By using \cite{Weng2013} the open circuit voltage can be substituted by the \ac{SoC} to obtain an analytical degradation map that can be discretized as shown in Fig.~\ref{fig:moura_deg}.

\subsection{Transformation to Individual Sized Battery Systems}
\label{sec:transsize}
The degradation maps presented in Sections \ref{sec:SIdegrad}~--~\ref{sec:empiricalDegFunction} refer to fixed cell charge capacities. To assess the impact for arbitrarily sized battery systems, the results from cell level need to be transferred to represent any battery system in size and power. In addition, it is important to transfer the maps from the current domain to the power domain to comply with the power system requirements. On cell level the degradation map from \ref{sec:SIdegrad}~--~\ref{sec:empiricalDegFunction} represents a map $ \dot{h}\left( Q^\mathrm{c}, I_\mathrm{bat}^{\mathrm{c}}\right)$ that describes the charge loss per time associated with the battery cell current $I_\mathrm{bat}^{\mathrm{c}}$ and the cell's absolute \ac{SoC} $Q^\mathrm{c}$. By multiplying this map with the cell's averaged open circuit potential $\bar{V}^\mathrm{c}_\mathrm{oc}$ the cell degradation ${J_\mathrm{deg}^\mathrm{c}}$ can be expressed in terms of energy per time as 
\begin{equation}
{J_\mathrm{deg}^\mathrm{c}} = \dot{h}\left( Q^\mathrm{c}, I_\mathrm{bat}^{\mathrm{c}}  \right) \bar{V}^\mathrm{c}_\mathrm{oc} \quad.
\label{eq:refDeg}
\end{equation}
\begin{figure}[!t]
	\centering
	\def\svgwidth{6cm}
	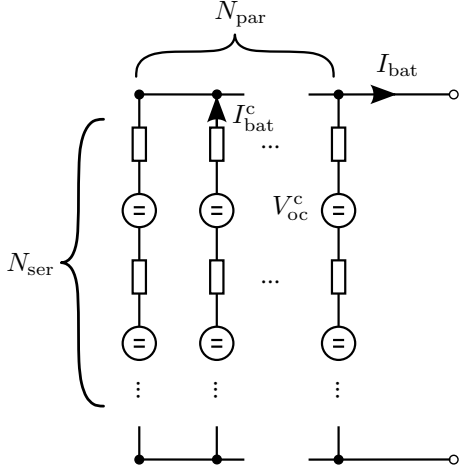
	\caption{Battery system consisting of $N_\mathrm{par}$ cell strings. Each string has $N_\mathrm{ser}$ cells connected in series.}
	\label{fig:batpack}
\end{figure}
As shown in Fig.~\ref{fig:batpack}, a battery system consists of $N_\mathrm{par}$ cell strings. Each string has $N_\mathrm{ser}$ cells connected in series. We can transform \eqref{eq:refDeg} to obtain the total capacity loss $J_\mathrm{deg}$ of an arbitrarily sized battery system in terms of energy 
\begin{equation}
J_\mathrm{deg} = N_\mathrm{par} \dot{h}\left( \frac{Q}{N_\mathrm{par}}, \frac{I_\mathrm{bat}}{N_\mathrm{par}} \right) N_\mathrm{ser} \bar{V}^\mathrm{c}_\mathrm{oc} \quad, \label{eq:totDeg}
\end{equation}
\noindent where $Q$ is the absolute state of charge of the individual sized battery. The energy capacity $C_\mathrm{E}$ of a battery system is 
$ C_\mathrm{E}=N_\mathrm{par}C_\mathrm{Q}^\mathrm{c}N_\mathrm{ser}\bar{V}^\mathrm{c}_\mathrm{oc}$. The total battery power is $P_\mathrm{bat}\approx I_\mathrm{bat} N_\mathrm{ser} \bar{V}^\mathrm{c}_\mathrm{oc}$, the absolute \ac{SoE} is $E \approx Q N_\mathrm{ser} \bar{V}_\mathrm{oc}$, and the normalized \ac{SoE} is defined as $E_\mathrm{n} = E/C_\mathrm{E}$. By substituting these definitions into \eqref{eq:totDeg}, we get
\begin{equation}
\frac{J_\mathrm{deg}}{C_\mathrm{E}} = \frac{\dot{h}\left(E_\mathrm{n} \  C_\mathrm{Q}^\mathrm{c},\frac{P_\mathrm{bat}}{C_\mathrm{E}}C_\mathrm{Q}^\mathrm{c}\right)}{C_\mathrm{Q}^\mathrm{c}} \quad,
\label{eq:totDegNorm}
\end{equation}
which is a normalized function that describes any battery system in size and battery power. This result shows us that the degradation map on cell level can be scaled by a linear coordinate transformation to any other battery system in size. The shape of the degradation map is invariant under scaling, which means that the transformation preserves the shape of the map. 

\subsection{Convexification of Degradation Maps}
\label{sec:convexhull}
Unfortunately, degradation maps are in general nonconvex (see Sections~\ref{sec:SIdegrad},\ref{sec:empiricalDegFunction}), such that efficient convex solvers cannot be applied for power optimization purposes. However, one can compute the convex hull of the degradation map \eqref{eq:totDegNorm} using Delaunay triangulation \cite{delaunay}. By evaluating the plane parameters $\vec{a}_1,\vec{a}_2,\vec{a}_3$ of the triangles from the convex hull, following piecewise affine mapping for the degradation $J_\mathrm{deg}$ is stated:
\begin{equation}
\frac{J_\mathrm{deg}}{C_\mathrm{E}} = \max\left(\vec{a}_1 \ \frac{P_\mathrm{bat}}{C_\mathrm{E}} + \vec{a}_2 \ E_\mathrm{n} + \vec{a}_3\right) \label{eq:pwadeg}
\end{equation}
To calculate the total capacity fade per time (kWh/h) of a battery with energy capacity $C_\mathrm{E}$ in dependence of the absolute \ac{SoE} $E$ and battery power $P_\mathrm{bat}$, one needs to transform \eqref{eq:pwadeg} to
\begin{equation}
J_\mathrm{deg} =  \max \left(\left[\vec{a}_1 \ \vec{a}_2 \ \vec{a}_3 \right]  \left[\begin{array}[h]{c}
P_{\mathrm{bat}} \\
E \\
C_\mathrm{E}
\end{array}\right]  \right)  \quad.
\label{eq:hull}
\end{equation}

\noindent Again this is a linear transformation, such that the shape of the map remains unchanged.

\subsection{Incorporation into an Optimal Control Framework}
In \cite{Fortenbacher2016_trans_operation,FortenbacherTrans2} it is shown, how the \ac{PWA} representation can be incorporated in an optimal control formulation and in an optimal battery sizing and placement problem. This can be achieved by incorporating an epigraph formulation of the \ac{PWA} map in the constraints.   
%\begin{align}
%&\min_{d} d \\
%& \text{s.t.} \nonumber \\
%&\vec{a}_1 P_\mathrm{bat} + \vec{a}_2 E + \vec{a}_3 C_\mathrm{E} \leq \vec{1} d \quad,
%\end{align}
\section{Example for Degradation Map Calculation}
\label{sec:example}
In this section we aim to show how degradation maps can be calculated using cycle test data. As explained in Section~\ref{sec:empiricalCycleTest}, we construct a pattern matrix that has caused the capacity loss for different test cycles. The authors in \cite{Wang2014} recorded the capacity loss for a LiMnNiCo/LiMn2O4 system for different \acp{DoD} and battery currents in an experimental setup. The charge capacity is $C_Q$ = 1.5 Ah. Table~\ref{tab:cycTest} lists the cycle test results of \cite{Wang2014} in terms of achieved full cycles, lost charge, and battery current. Also the resulting pattern numbers $p$ and time intervals $T_b$ calculated with \eqref{eq:p} and \eqref{eq:Tb} are shown. Since the tests have different discretization levels, we need to define a non-uniform grid for $\vec{I}_\mathrm{s}$.  
\begin{table}[!t]
	\caption{Cycle test results from \cite{Wang2014} and parameters for pattern reconstruction matrix \vec{M} and $T_\mathrm{b}$.}
	\label{tab:cycTest}
	\centering
	\begin{tabular}{llcrcrrrr@{}}
		\toprule
	 \# &	$I_\mathrm{bat}$ (A) &$T_{\mathrm{b}}$ (h) & $n_\mathrm{cyc}$ & DoD & \multicolumn{2}{c}{$p$}  & \multicolumn{2}{c}{$Q_\mathrm{s}$ (Ah)} \\
		\midrule
		&					 &					& 3333	   	& 0.1	&$p_{11}$ & 66.6e3 & $Q_{11}$ & 0.33 \\
		&					 &					& 3067	   	& 0.2	&$p_{12}$ & 20.4e3 & $Q_{12}$ & 0.45\\
	1	&	5.25			 &   0.057			& 2500	   	& 0.5	&$p_{13}$ & 10.0e3 & $Q_{13}$ & 0.45\\
		&					 &					& 2000	   	& 0.7	&$p_{14}$ & 5.7e3  & $Q_{14}$ & 0.45\\
		&					 &					& 666	   	& 0.9	&$p_{15}$ & 1.4e3  & $Q_{15}$ & 0.18\\				  
		\midrule
		&					 &					& 3333	   	& 0.3	&$p_{21}$ & 2.22e4 & $Q_{21}$ & 0.40\\
	2	&	3				 &	0.160			& 2800	   	& 0.5	&$p_{22}$ & 11.2e3 & $Q_{22}$ & 0.45\\
		&					 &					& 2500	   	& 0.7	&$p_{23}$ & 6.6e3& $Q_{23}$ & 0.45\\	
		\bottomrule
	\end{tabular}
\end{table}
For the test \#1 we need to define $n_\mathrm{bd} = 5$ SoC bands (0.10,0.30,0.50,0.70,0.90). The corresponding pattern matrix $\vec{M}_{\mathrm{p},1}$ and right hand side vector $\vec{Q}_{\mathrm{s},1}$ is constructed as
\begin{equation}
\vec{M}_{\mathrm{p},1} = T_{\mathrm{b},1}\left[\arraycolsep=1.5pt\begin{array}{ccccc} 0 & 0 & p_{11} & 0 & 0 \\ 0 & p_{12} & p_{12} & 0 & 0 \\
0 &p_{13} & p_{13} & p_{13} & 0 \\ p_{14} &p_{14} & p_{14} & p_{14} & 0 \\ p_{15} &p_{15} & p_{15} & p_{15} & p_{15}
\end{array} \right] \ \vec{Q}_{\mathrm{s},1} = \left[\arraycolsep=1.5pt\begin{array}{c} Q_{11} \\ Q_{12} \\ Q_{13} \\ Q_{14} \\ Q_{15} \end{array} \right] .
\end{equation}

\noindent and for test \# 2 $n_\mathrm{bd} = 3$ with SoC bands = (0.16,0.50,0.83) and the corresponding pattern matrix and right hand side vector
\begin{equation}
\vec{M}_{\mathrm{p},2} = T_{\mathrm{b},2}\left[\begin{array}{ccc} 0 & p_{21} & 0 \\ p_{22} & p_{22} & 0 \\ p_{23} & p_{23} & p_{23} \\ 
\end{array} \right] \ \vec{Q}_{\mathrm{s},2} = \left[\begin{array}{c} Q_{21} \\ Q_{22} \\ Q_{23}  \end{array} \right] .
\end{equation}

\begin{figure}[!t]
	\centering
	\includegraphics[width=\columnwidth]{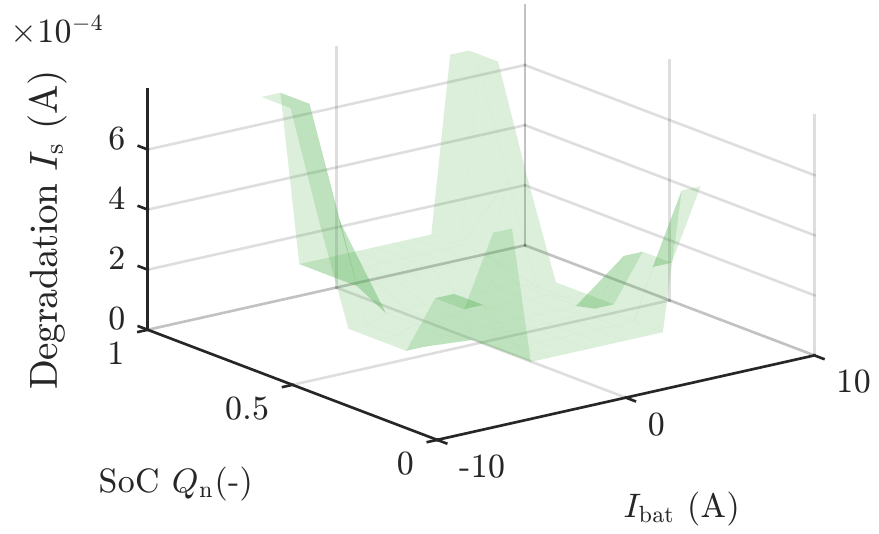}
	\caption{Identified degradation map of LiMnNiCo/LiMn2O4 battery cell using cycle test data from \cite{Wang2014}.}
	\label{fig:wang_deg}
\end{figure}

We arrange the matrices according to \eqref{eq:pattern} and solve \eqref{eq:Is} to obtain the degradation map that is depicted in Fig.~\ref{fig:wang_deg}. 

\section{Degradation Maps}
\label{app:deg}
\begin{figure}[!t]
	\center
	\includegraphics[width=9cm]{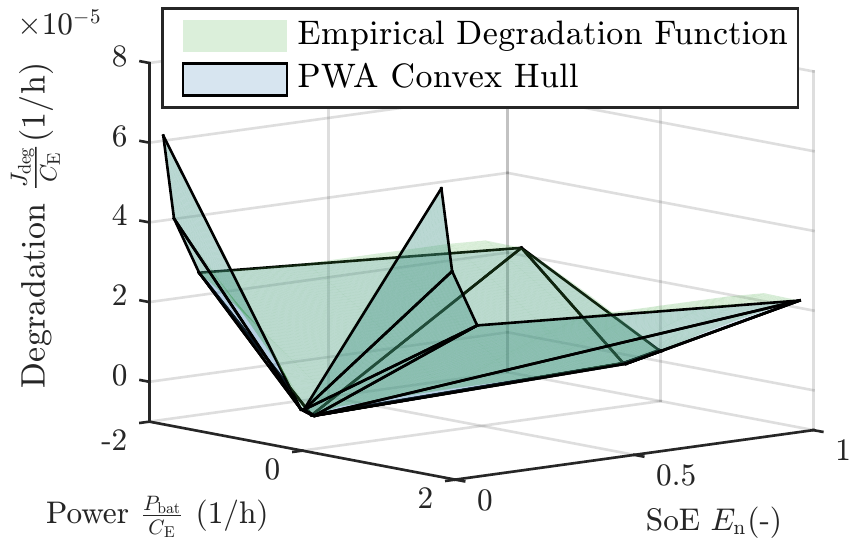}
	\caption{Normalized degradation map using the data of \cite{Forman2012} for a LiFePO4 based cathode chemistry. The blue surface is the piecewise affine (PWA) convex hull \eqref{eq:hull} of the normalized representation of \eqref{eq:totDegNorm} (green surface).}
	\label{fig:degMoura}
\end{figure}

\begin{figure}[!t]
	\center
	\includegraphics[width=9cm]{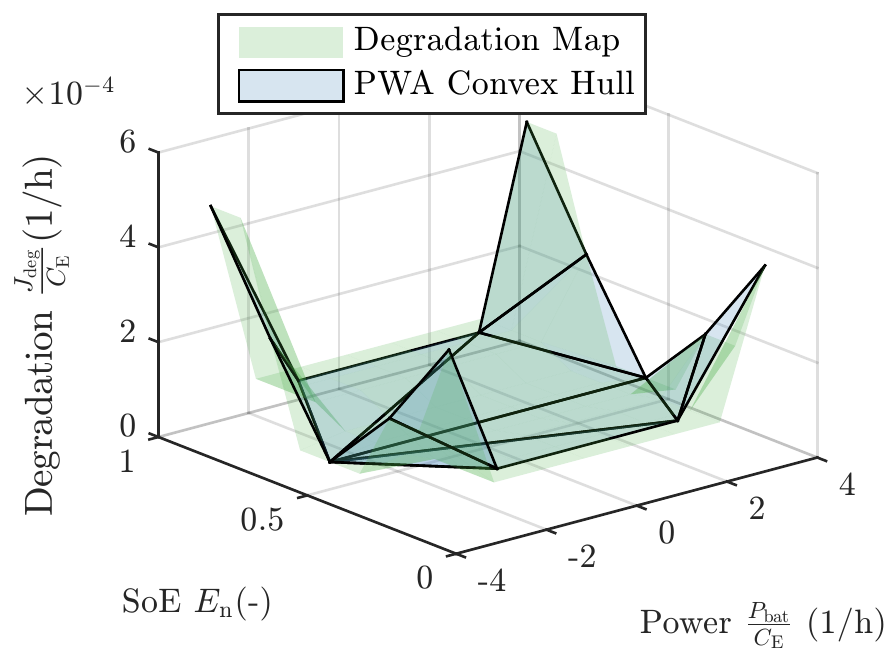}
	\caption{Normalized degradation map for a LiMnNiCo/LiMn2O4 based cathode chemistry using the degradation map results from Section \ref{sec:example}. The blue surface is the piecewise affine (PWA) convex hull \eqref{eq:hull} of the normalized representation of \eqref{eq:totDegNorm} (green surface).}
	\label{fig:degWang}
\end{figure}

\begin{figure}[!t]
	\center
	\includegraphics[width=9cm]{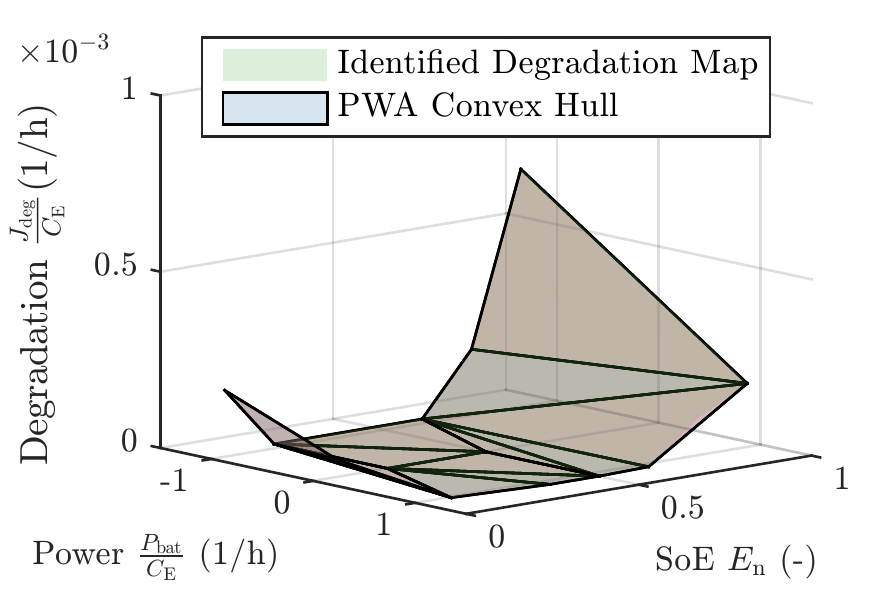}
	\caption{Normalized degradation map for the DUALFOIL \cite{Fuller1994} cell comprising a LiCoO2 based cathode chemistry using the degradation map results from \cite{fortenbacherPSCC14}. The blue surface is the piecewise affine (PWA) convex hull \eqref{eq:hull} of the normalized representation of \eqref{eq:totDegNorm} (green surface).}
	\label{fig:degDualFoil}
\end{figure}

Next, we present the degradation maps and their \ac{PWA} approximations in the Figures~\ref{fig:degMoura},\ref{fig:degWang}, and \ref{fig:degDualFoil} for the proposed technologies. The obtained degradation maps on cell level are normalized as explained in Section \ref{sec:transsize} and the convex hulls of the normalized maps are calculated as described in Section \ref{sec:convexhull}. The corresponding plane parameters of the convex \ac{PWA} approximations are listed in the Tables~\ref{tab:degmap_lifepo},\ref{tab:degmap_wang}, and \ref{tab:degmap_lico} in the Appendix \ref{app:tables}. Comparing the different degradation maps, we can see that the battery current for LFP (Fig.~\ref{fig:degMoura}) and NMC/LMO systems (Fig.~\ref{fig:degWang}) is the driving factor for battery degradation, while the degradation of the LCO system (Fig.~\ref{fig:degDualFoil}) is more sensitive to the SoE.  

We also compute the root-mean squared errors (RMSEs) and normalzized root-mean squared errors (NRMSEs) of the \ac{PWA} approximation for each battery technology. The low NRMSEs (<3\%) indicate that the error of the convex hull approximation related to the original degradation map is rather small. This means that the original maps are almost convex, such that the \ac{PWA} representation is an appropriate proxy for battery degradation.  

\begin{table}[!t]
	\caption{Root-Mean Squared Errors (RMSEs) and normalized Root-Mean Squared Errors (NRMSEs) of PWA approximation.}
	\label{tab:RMSE}
	\centering
	\begin{tabular}{llcc}
		\toprule
		Battery Technology & Cathode & RMSE & NRMSE \\
						   & Material &	(1/h)  &  (\%)\\
		\midrule
		LCO &LiCoO2 & 7.42e-6 &  1.06 \\
		NMC/LMO &LiMnNiCo/& 5.37e-6 & 1.07 \\
				& LiMn2O4 \\
		LFP & LiFePO4 & 2.01e-6 & 3.33 \\
		\bottomrule
	\end{tabular}
\end{table}

\section{Conclusion}
In this paper we have presented a novel method to efficiently model battery degradation in power system applications. We use the concept of degradation maps that associate individual discrete control actions with an incremental capacity loss. In contrast to calendar and cylce life models, degradation maps allow to characterize battery degradation for arbitrary usage pattern. We have shown that degradation maps can be scaled to represent any battery system in size and power and their convex hull representations are appropriate approximations. In this way, we are able to efficiently evaluate battery degradation in power system simulations and to incorporate battery degradation in optimal control frameworks. With the published degradation parameters, we think that we can provide a meaningful and transparent database for other researchers to benchmark different battery technologies in a common framework.

\appendices
\section{Degradation Map Parameters}
\label{app:tables}
\begin{table}[!h]
	\centering
	\caption{Degradation map for the LiFePO4 battery from \cite{Forman2012}.}
	\begin{tabular}{rrr} 
		\toprule
		\multicolumn{1}{c}{$\vec{a}_1$ (-)} & \multicolumn{1}{c}{$\vec{a}_2$ (1/h)} &   \multicolumn{1}{c}{$\vec{a}_3$ (1/h)} \\ 
		\midrule
		-3.452e-05 & -7.058e-04 & -3.291e-07\\
		-2.620e-05 & -2.067e-04 & -1.763e-07\\
		-1.595e-05 & -5.485e-06 & -1.657e-06\\
		-1.811e-05 & -6.110e-05 & -2.774e-08\\
		-1.162e-05 & 2.548e-06  & -1.818e-06\\
		-1.064e-05 & 2.010e-05  & -1.760e-05\\
		0.000e+00  & -6.110e-05 & 3.049e-07\\
		0.000e+00  & -6.110e-05 & 3.049e-07\\
		0.000e+00  &2.548e-06   & -1.605e-06\\
		0.000e+00  & 2.010e-05  & -1.740e-05\\
		0.000e+00  & 2.548e-06  & -1.605e-06\\
		0.000e+00  & 2.010e-05  & -1.740e-05\\
		1.811e-05  & -6.110e-05 & -2.774e-08\\
		3.452e-05  &-7.058e-04  & -3.291e-07\\
		2.620e-05  & -2.067e-04 & -1.763e-07\\
		1.162e-05  & 2.548e-06  & -1.818e-06\\
		1.595e-05  & -5.485e-06 & -1.657e-06\\
		1.064e-05  & 2.010e-05  & -1.760e-05\\
		\bottomrule
	\end{tabular}
	\label{tab:degmap_lifepo}
\end{table}

\begin{table}[!h]
	\centering
	\caption{Degradation map for the LiMnNiCo/LiMn2O4 battery from \cite{Wang2014}.}
	\begin{tabular}{rrr} 
		\toprule
		\multicolumn{1}{c}{$\vec{a}_1$ (-)} & \multicolumn{1}{c}{$\vec{a}_2$ (1/h)} &   \multicolumn{1}{c}{$\vec{a}_3$ (1/h)} \\ 
		\midrule
		-1.608e-04 & -9.698e-04 & -7.274e-05\\ 
		-1.373e-04 & -7.065e-04 & -6.940e-05\\
		-1.998e-04 & 1.055e-03 & -1.169e-03\\
		0.000e+00 & 1.549e-04 & -1.975e-05\\
		0.000e+00 & -9.016e-05 & 1.027e-04\\
		0.000e+00 & -9.016e-05 & 1.027e-04\\
		0.000e+00 & 1.549e-04 & -1.975e-05\\
		-2.083e-4 & 1.150e-03 & -1.265e-03\\
		1.608e-4 & -9.698e-04 & -7.274e-05\\
		1.373e-4 & -7.065e-04 & -6.940e-05\\
		1.998e-4 & 1.055e-03 & -1.169e-03\\
		2.083e-4 & 1.150e-03 & -1.265e-03\\
		\bottomrule
	\end{tabular}
	\label{tab:degmap_wang}
	\vspace{-0.1cm}
\end{table}

\begin{table}[!h]
	\centering
	\caption{Degradation map for the DUALFOIL \cite{Fuller1994} LiCoO2 battery.}
	\begin{tabular}{rrr} 
		\toprule
		\multicolumn{1}{c}{$\vec{a}_1$ (-)} & \multicolumn{1}{c}{$\vec{a}_2$ (1/h)} &   \multicolumn{1}{c}{$\vec{a}_3$ (1/h)} \\ 
		\midrule
		-1.156e-04 &-1.231e-03 & 1.354e-04 \\
		-1.262e-07 & 3.849e-08 &-1.826e-08 \\
		-1.162e-07 &-6.953e-05 &1.490e-05 \\
		-1.162e-07 &-1.893e-10 &1.081e-09 \\
		-3.392e-05 &-5.953e-04 &9.002e-05 \\
		-7.040e-05 &1.220e-03 &-8.624e-04 \\
		7.582e-08 & 1.610e-06 &-8.038e-07 \\
		1.299e-06 &2.063e-05 &-1.167e-05 \\
		1.299e-06 &6.624e-04 &-4.243e-04 \\
		-2.114e-04 &3.413e-03 &-2.742e-03 \\
		7.582e-08 &-1.893e-10 & 1.081e-09 \\
		4.507e-06& -3.357e-05 & 7.194e-06 \\
		3.081e-07& -9.033e-07 & 1.946e-07 \\
		\bottomrule
	\end{tabular}
	\label{tab:degmap_lico}
\end{table}

\bibliographystyle{IEEEtran}
\bibliography{literature}

\end{document}